\documentstyle[prb,subeqn,aps,psfig]{revtex}

\begin{document}
\draft
\title{\bf Magnetic field induced control of breather dynamics in a single 
plaquette of Josephson junctions.}

\author{ M. V. Fistul, S. Flach and A. Benabdallah}
\address{Max-Planck-Institut f\"ur Physik komplexer Systeme,
N\"othnitzer Stra\ss e 38, D-01187 Dresden, Germany}

\date{\today}

\wideabs{ 

\maketitle

\begin{abstract}
We present a theoretical study of {\it inhomogeneous} dynamic (resistive) 
states in a single plaquette consisting of three Josephson junctions.
Resonant interactions of such a breather state with electromagnetic 
oscillations manifest themselves by
resonant current steps and voltage jumps in the current-voltage characteristics. 
An externally applied magnetic field leads to a variation of
the relative shift between the Josephson current oscillations of 
two resistive junctions.  By making use of the rotation wave approximation 
analysis and direct numerical simulations we show that this effect allows 
to effectively control the breather instabilities, e. g. 
to increase (decrease) the height of the resonant steps and to suppress the 
voltage jumps in the current-voltage characteristics. 
\end{abstract}
}
Discrete Josephson coupled systems have attracted recently 
a lot of attention due to the prediction \cite{Floria96,Spicci99,mto99} 
and the following observation 
of {\it intrinsic dynamic localized } states.
\cite{Binder00,bau00,tmo00,tmbo00} 
These breather states appear in homogeneously dc current driven 
Josephson junction ladders and arrays in the form of various 
{\it spatially inhomogeneous} voltage patterns. 
These voltage patterns are characterized by a few junctions being 
in the resistive (whirling) state while the rest of all junctions reside in the 
superconducting (libration) states. The breather states manifest themselves 
through 
additional branches in the current-voltage ($I$-$V$) 
characteristics of a system \cite{Binder00,bau00,tmo00,tmbo00} and, 
moreover, can be directly visualized by 
using the low temperature scanning microscopy (LTSM) technique.
\cite{Binder00,bau00} 
Note here, that the origin of such a dynamic localization is not the 
presence of disorder but a peculiar interplay between nonlinearity and 
discreteness \cite{fw98}, and therefore, breather states can play the same important 
role for underdamped Josephson junction arrays as Josephson vortices do for  
parallel arrays. \cite{UstinovRev}

The important peculiarity of inhomogeneous resistive states is a  
strong resonant interaction with linear electromagnetic excitations, 
e. g. electromagnetic waves or localized electromagnetic oscillations. 
It was predicted in Ref. \onlinecite{tmbo00} and carefully studied in Ref.
\onlinecite{MFFZP,bff} that the resonant interaction of 
the breather state with  
electromagnetic excitations leads to  
resonant steps (a sharp current increase) and various switching phenomena 
between different resistive states 
(voltage jumps) in the $I$-$V$ curves. 

The simplest case where inhomogeneous resistive states 
can be obtained,\cite{mto99,bff} is a dc current driven single anisotropic 
plaquette containing three Josephson junctions, as presented in Fig. 1. 
It consists of two {\it vertical} junctions parallel to the bias current 
$\gamma$ direction, and a {\it horizontal} junction in the 
transverse direction.
The dynamics of the system crucially depends on two parameters:
the anisotropy $\eta=\frac{I_{cH}}{I_{cV}}$, where 
$I_{cH}$ and $I_{cV}$ are respectively the critical currents of horizontal 
and vertical junctions, and the discreteness parameter 
(normalized inductance of the cell), $\beta_L$.

\begin{figure}[!hbp]
\vskip 0.6cm

\centerline{\psfig{figure=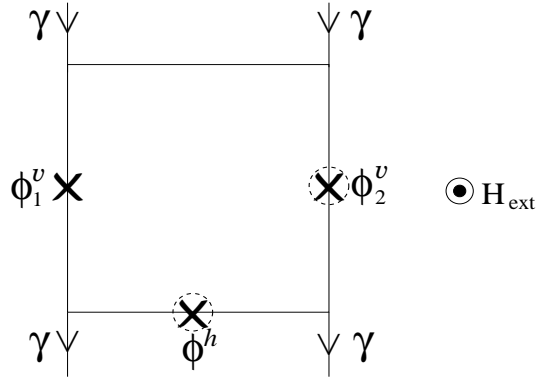,angle=-90,width=7cm,height=5cm}}
\caption{Sketch of the plaquette with three Josephson junctions 
(marked by crosses) in the presence of uniform dc bias $\gamma$ 
and an externally applied magnetic field $H_{ext}$. Dashed circles denote 
junctions in the resistive (whirling) state.}
\end{figure}

In Ref. \onlinecite{bff} we have shown that inhomogeneous resistive 
states can be observed in this system as the anisotropy $\eta$ is not large.
These two breather states are characterized by only one (left or right) 
vertical junction and 
the horizontal junction being resistive. 
We found also various further features of the states, 
e.g. resonant current steps and voltage jumps, 
in the $I$-$V$ curves. These features can be considered 
as "fingerprints" of 
the resonant interaction of the breather state with 
electromagnetic oscillations (EOs). The voltage positions of 
the resonances  depend strongly on
the parameters $\eta$ and $\beta_L$. 

In this Letter, we report on the properties of breather states or even 
more specifically, on the 
resonant interaction of the breather states with EOs in 
the presence of an externally applied magnetic field. 
We show that the 
magnetic field allows to control the {\it strength} of the resonant 
interaction  
just by a fine tuning of the relative phase shift between the 
Josephson current oscillations of the two resistive junctions.

In order to quantitatively characterize the magnetic field induced control we 
write the set of equations for the time-dependent Josephson phases 
$\phi_1^v$, $\phi_2^v$ and $\phi^h$:
\begin{eqnarray}
\nonumber
{\cal N} (\phi^{v}_1) &=& \gamma - \frac{1}{\beta_L}
(\phi^{v}_1 - \phi^{v}_2 + \phi^{h} + 2 \pi f) \; , \\ \label{phi1}
{\cal N} (\phi^{v}_2) &=& \gamma + \frac{1}{\beta_L} 
(\phi^{v}_1 - \phi^{v}_2 + \phi^{h} + 2 \pi f) \; , \\ \nonumber
{\cal N} (\phi^{h}) &=& \frac{1}{\eta \beta_L}
(\phi^{v}_1 - \phi^{v}_2 + \phi^{h} + 2 \pi f) \; ,
\end{eqnarray}
where the nonlinear operator ${\cal N}(\phi) = \ddot \phi + 
\alpha \dot \phi + \sin (\phi)$, and the dissipation parameter 
$\alpha$ determines the effective damping in the system. 
Here, we use the inverse of the plasma frequency $\omega_J$ as 
the unit of time. The externally applied magnetic field $H_{ext}$ is characterized by 
the frustration $f=\frac{\Phi_{ext}}{\Phi_0}$, i. e. 
the magnetic flux threading the cell normalized to the magnetic flux quantum.  
These dynamic equations have been derived in Ref. \onlinecite{bff}, using 
the resistively shunted model for Josephson junctions, the Kirchhoff's current
laws and the flux quantization law.

The Josephson phases are naturally decomposed as follows 
(below we consider a particular breather case as the left vertical 
junction is in the superconducting state):
\begin{eqnarray}
\nonumber
\phi^{v}_1 (t) &\approx&  c_1+\delta_1^v(t)~~, \\ \label{phiapprox}
\phi^{v}_2 (t) &\approx& \Omega t + c_2+\delta_2^v(t)~~, \\ \nonumber
\phi^{h} (t) &\approx& \Omega t +\delta^h(t)\, ,
\end{eqnarray} 
where the breather frequency $\Omega$ and the phase shifts $c_1$ and $c_2$ 
are obtained using a {\it dc analysis},\cite{bff} i. e. 
neglecting the Josephson phase oscillations $\delta (t) $:
\begin{eqnarray}
\label{cur_fre}
\nonumber
\Omega &=&  \frac{\gamma}{\alpha(1+\eta)}~~, \\ \label{cun}
c_1 &=& {\rm asin} \left ( \frac{1+2\eta}{1+\eta} \gamma \right )~~, \\ 
\label{2-9c}
\nonumber 
c_2 &=& c_1 +
\frac{\beta_L \gamma \eta}{1+\eta} + 2\pi f~~.
\end{eqnarray}
In Eq. (\ref{phiapprox}) $\delta^v_{1,2}(t),\, \delta^{h}(t)$ correspond to EOs
excited in the presence of the breather state.   
By making use of the linearization of 
Eq. (\ref{phi1}) around the breather state (\ref{phiapprox}) we derive two 
characteristic frequencies of EOs:
\begin{eqnarray}
\label{omega34}
\nonumber
|\omega_{\pm}| = \sqrt{ F \pm \sqrt{ F^2-G}} \;\;, ~~~~~~~~~~~~~~\\
F~=~\frac{1}{2}\cos(c_1) + \frac {1 + 2\eta}{2\eta\beta_L}
\;\;,~~ 
G~=\cos(c_1) \frac{(1 + \eta) }{\eta\beta_L}
\;\;. \label{3-G}
\end{eqnarray}

Thus, as the breather frequency $\Omega$ matches any of the EOs frequencies,
namely $\Omega=\omega_{\pm}$, a {\it primary} resonance appears, and 
the $I$-$V$ curve displays a resonant current step. Using the 
rotation wave approximation method 
elaborated in \cite{Kulik} we find that the  strength 
of the primary resonance is determined by the {\it difference} of two Josephson 
current oscillations: $ \sin (\Omega t+c_2)-\sin (\Omega t)$. It leads to 
the magnetic field dependent magnitude of the resonant step $\Delta \gamma$ as
\begin{equation}\label{ResStep}
\Delta \gamma ~\propto~\left |\sin \frac{c_2}{2} \right | ~=~  
\left |{\sin\left( \frac{c_1}{2} +\frac{\beta_L \gamma \eta}{2(1+\eta)}
 +\pi f \right)} \right |\,.
\end{equation}

Next, we turn to the {\it parametric} resonant interaction of the breather state 
with EOs,
as $\Omega~=2\omega_{\pm}$. In order to analyze this case 
we introduce the new variables 
$D(t)~=~\delta_2^v(t)-\delta^h(t)$ and $S(t)~=\delta_2^v(t)+\delta^h(t)$.
The equations for $D(t)$ and $S(t)$ are given by:

\begin{eqnarray}
  \nonumber
  \label{beta1}
  \ddot D + \alpha \dot D + 
  \omega_{\pm}^2 D 
  &=&
  -D h_1(t)+S h_2(t)~~, 
  \\ \label{beta2}
  \ddot S+ \alpha \dot S 
  + \frac{\eta-1}{\eta+1}\omega_{\pm}^2 D&=&
 -S h_1(t)+D h_2(t)~~,
  \end{eqnarray}
where the time dependent coefficients $h_1(t)~=~\cos \frac{c_2}{2} 
\cos (\Omega t)$ and $h_2(t)~=~\sin \frac{c_2}{2} \sin (\Omega t)$.
Note here that we neglected all nonresonant terms in Eq. (\ref{beta2}). 
\cite{comment}

Similarly to a well known case of a parametrically driven 
harmonic oscillator \cite{Landau}
we find that in the dissipative case with two degrees of freedom 
(Eq. (\ref{beta2}) ) the breather state can become parametrically unstable, 
when the Josephson phases 
$D(t),S(t)~\simeq~e^{\lambda t}$ with $\lambda~>~0$. This instability 
depends on the effective amplitude of the parametric 
Josephson current oscillations $h_{eff}(f)$ , and is determined by 
the condition:
\begin{equation}\label{ParInst}
h_{eff}(f)~=~\sqrt{(1-\eta)^2 + 4\eta \cos^2\left( \frac {c_2}{2}
  \right )}~>~\alpha \Omega
\end{equation}
Moreover, the strength of the 
{\it combination resonance}, as $\Omega~=~\omega_+ \pm \omega_-$, between 
the breather state and EOs is also determined by the parameter $h_{eff}(f)$. 
The condition of the combination resonance instability is 
\begin{equation}\label{CombInst}
h_{eff}(f)~>~\alpha \sqrt{\omega_+ \omega_-}~~.
\end{equation}
Thus, the applied magnetic field allows to control the strength of the 
parametric and combination resonances through the magnetic field dependent 
phase shift between two Josephson current oscillations, $c_2$. 

To test the obtained analysis and to show how the applied magnetic field 
changes 
the $I$-$V$ curve we perform direct numerical simulations of  
Eq. (\ref{phiapprox}). Details of the numerical procedure were 
given elsewhere.\cite{bff} To establish a large dc current bias region where 
various types of resonances occur we use the discreteness 
parameter $\beta_L~=~1$ and the anisotropy $\eta~=~0.5$.  
We decreased (increased) the dc bias current
and calculated the dc voltage $V~=~<\dot \phi_2^v>$.
Numerically simulated $I$-$V$ curves for different values of 
the applied magnetic field are shown in Fig. 2.

In the absence of the magnetic field ( see Fig. 2a) 
we observed two resonant steps
labeled as $A$ and $B$, and the switching phenomenon 
(the voltage jump to the homogeneous resistive state) labeled as $C$.
All types of resonant interactions of the breather state with EOs 
are present in this $I$-$V$ curve:
the parametric resonance $A$ leads to the the resonant step 
at upper values of the dc bias $\gamma$, the primary resonance $B$ leads to the 
resonant step at lower values of the dc bias $\gamma$, 
and the voltage jump $C$ is the consequence of the combination resonance. 
\cite{bff}
 \begin{figure}
\centerline{
\psfig{figure=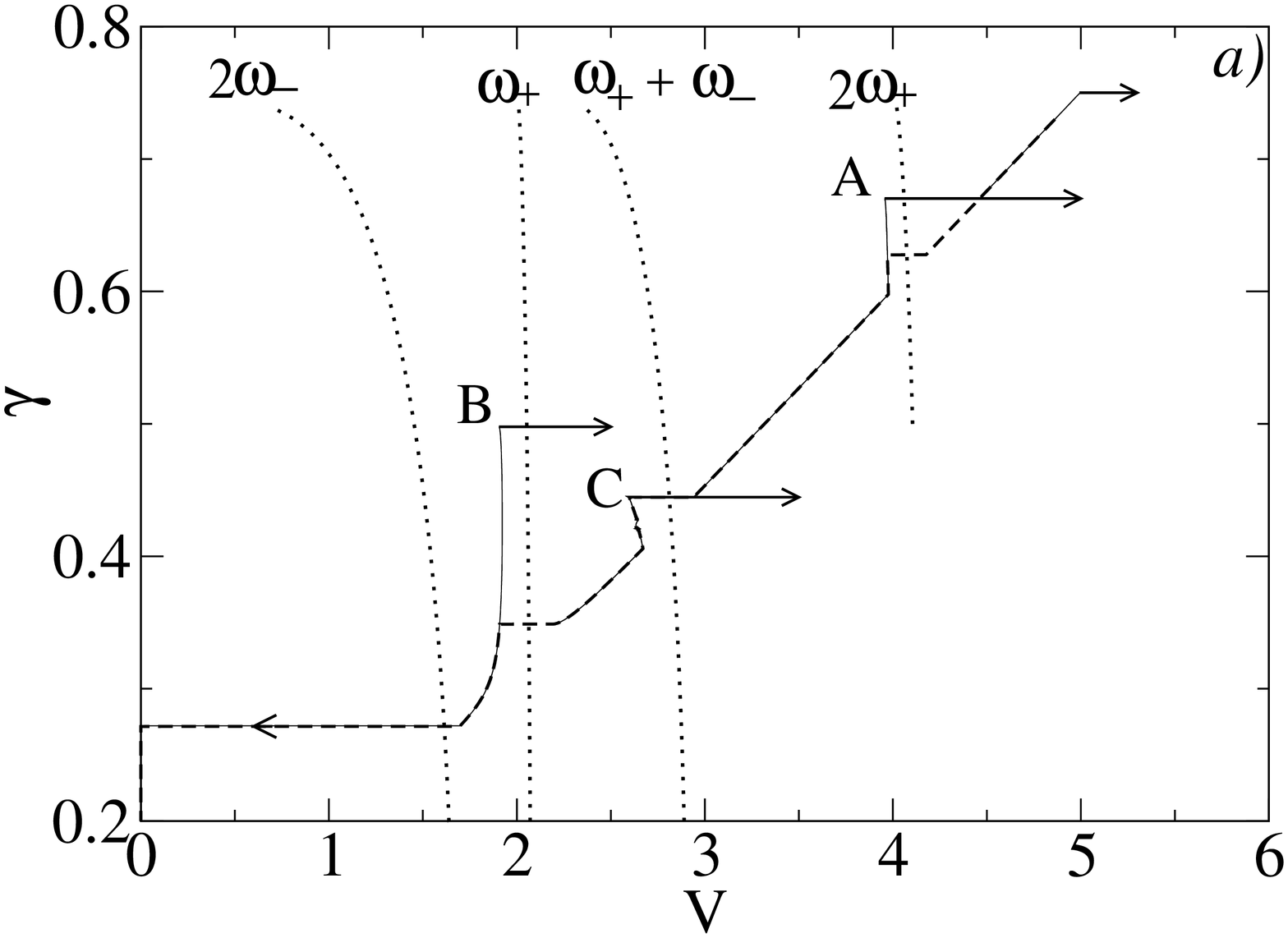,angle=-90,width=8cm,height=6cm}}
\centerline{\psfig{figure=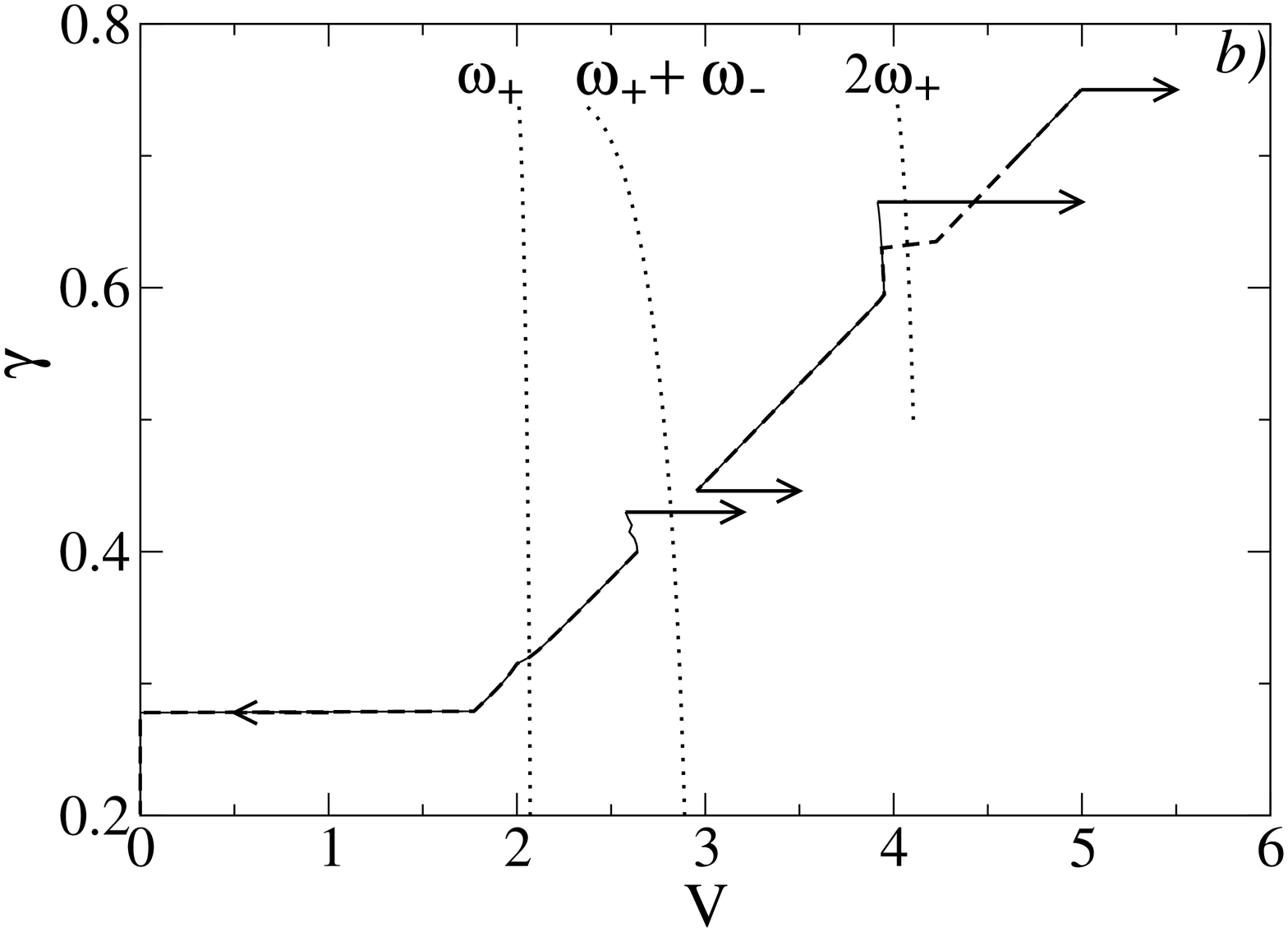,angle=-90,width=8cm,height=6cm}}
\centerline{\psfig{figure=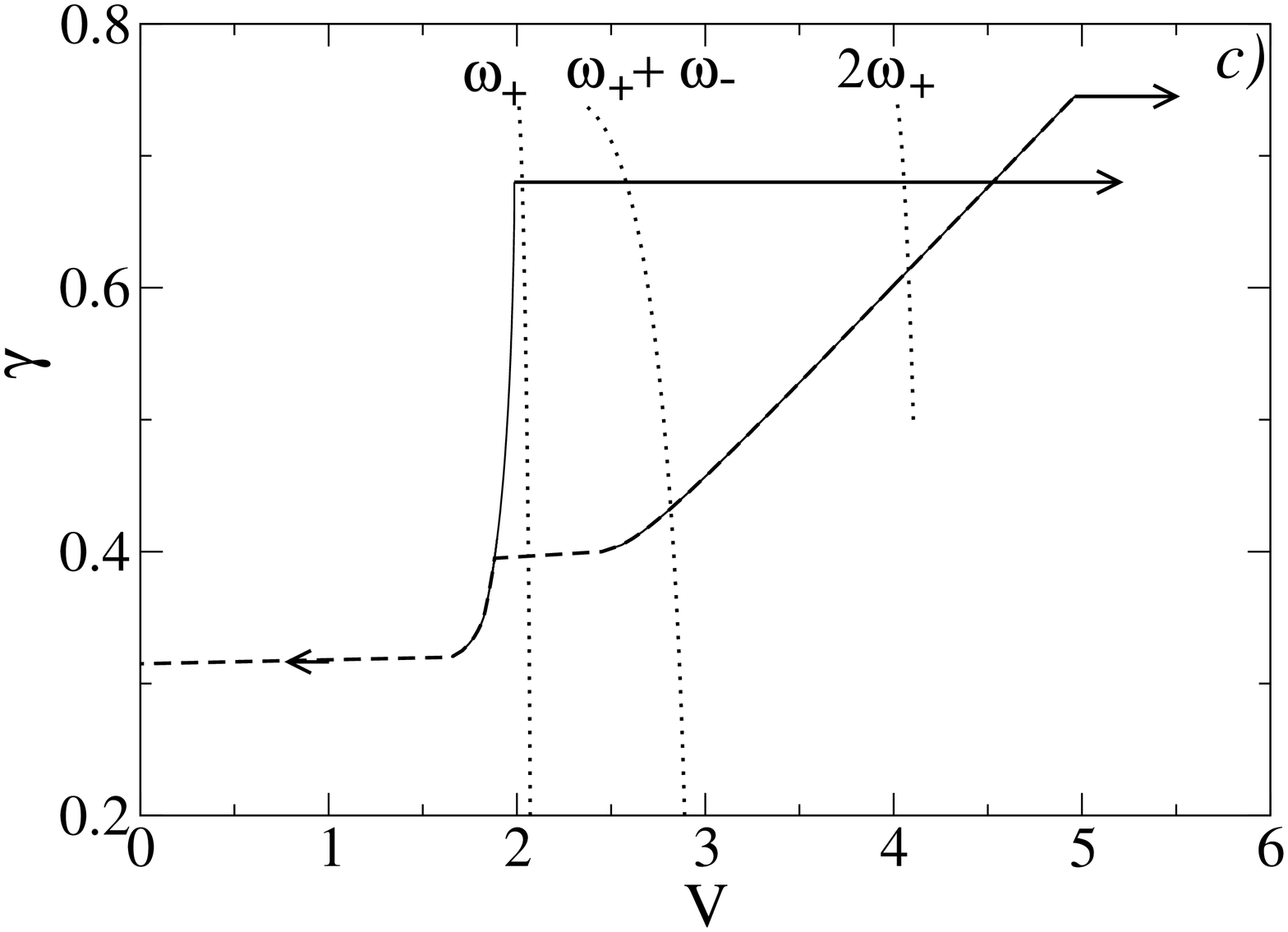,angle=-90,width=8cm,height=6cm}}
\caption{$I$-$V$ characteristics of a breather state for $\beta_L=1,$ 
$\alpha =0.1$ and $\eta=0.5$. The values of the magnetic field  are (a) $f=0$;
(b) $f=-0.1$; (c) $f=0.4$. The dotted lines show the dependence of the 
characteristic frequency combinations of EOs on the dc bias $\gamma$ 
(Eq. (\ref{omega34}) ). Arrows indicate the various switching processes.}
\end{figure}
In the presence of an applied magnetic field we find that the $I$-$V$ curve 
changes drastically. Thus, for a small magnetic field $f~=~-0.1$ 
the lower voltage resonant step $B$ practically disappears, but the 
switching phenomenon $C$ becomes even stronger (Fig. 2b).
In the opposite limit of a large magnetic field, as $f=0.4$, the 
parametric 
and the combination resonances ($A$, $C$)
disappear from the $I$-$V$ curve but the magnitude 
of the lower voltage 
resonant step $B$ (primary resonance) increases (Fig. 2c).

The dependence of the magnitude of the lower resonant step on the frustration 
displays a minimum on small negative values of $f$ and reaches a maximum as 
$f~\simeq~0.4$ (circles, in Fig. 3a). 
We find that the outcome of our numerical simulation 
is in good accord with the 
theoretical analysis (Eq. (\ref{ResStep}) and solid line in Fig. 3a).
\begin{figure}
\centerline{\psfig{figure=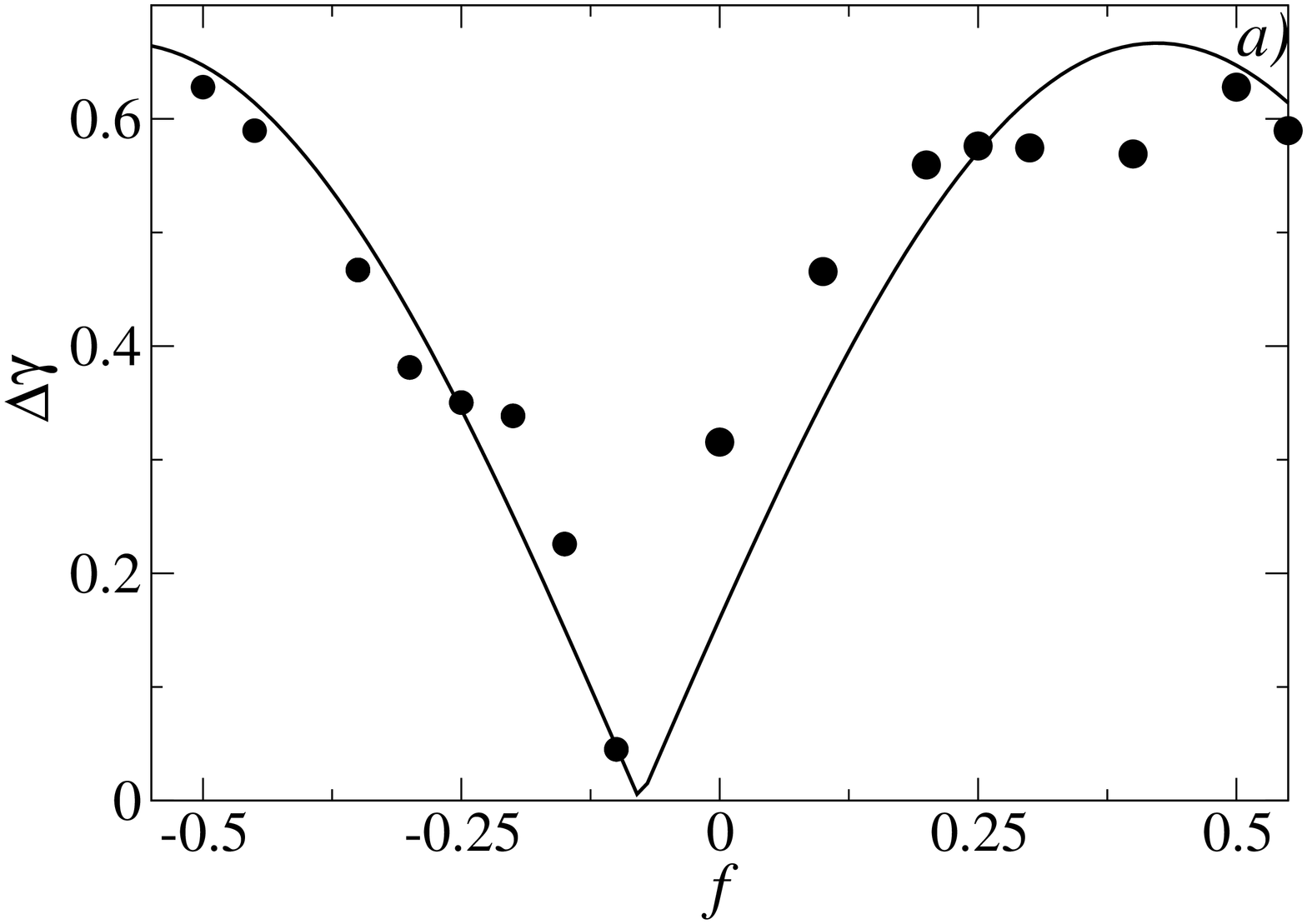,angle=-90,width=8cm,height=6cm}}
\centerline{\psfig{figure=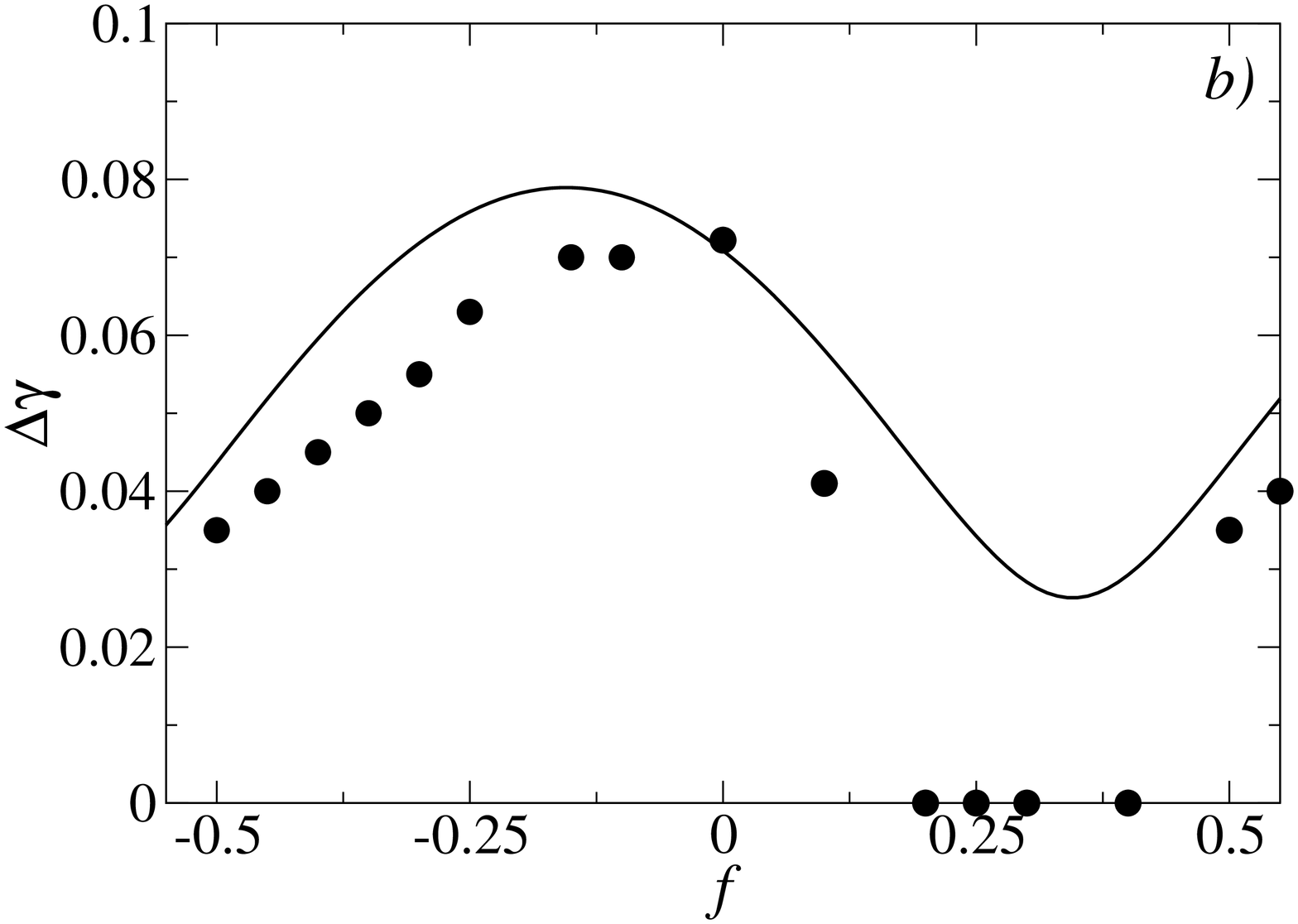,angle=-90,width=8cm,height=6cm}}
\caption{The dependence of the magnitude of the resonant steps on 
the frustration $f$: (a) the lower resonant step (primary resonance, $B$). 
The numerical results (circles) and the analytical prediction 
(Eq. (\ref{ResStep}) ) are shown correspondingly by circles and solid line.
(b) the upper resonant step (parametric resonance, $A$), 
the numerical results (circles) and the analytical prediction (solid line) 
based on the expression for $h_{eff}(f)$ (Eq. (\ref{ParInst}) ) 
are shown. 
}
\end{figure}       

The numerically simulated dependence of the magnitude of the 
upper resonant step $A$ is shown in Fig. 3b. 
This dependence displays a peculiar oscillation and is 
in qualitative agreement with the magnetic field dependence of $h_{eff}(f)$. 
Moreover, 
as suggested by the inequality in  Eq. (\ref{ParInst}) 
is valid, 
in the region of frustration $0.2~<f~<0.37$ a specific 
"window" of magnetic field occurs.
In this region of the magnetic field  the parametric 
resonant interaction is very weak and the resonant step $A$ disappears 
from the $I$-$V$ curve. Note here that the width of the "windows" 
increases with the 
damping parameter $\alpha$ (Eq. (\ref{ParInst}) ), 
and this allows for an accurate 
measurement of damping at low temperatures as $\alpha$ is very small.

In the same region of 
frustration, a similar "window" exists for the combination resonance and 
correspondingly the switching of the breather state to the 
homogeneous resistive state is suppressed. 
This effect has a peculiar consequence. Because of the instability of 
the breather state and the corresponding switching to the 
homogeneous resistive state ( see Fig. 2b), it can be very hard 
to experimentally observe the lower part of the 
$I$-$V$ curve. However, in the presence of a magnetic field the breather can 
pass this region of dc bias current and penetrate to the "forbidden" current 
region.

In conclusion we presented an analysis and direct numerical simulations of 
the inhomogeneous resistive state in a single plaquette consisting of three 
Josephson junctions. We find that the $I$-$V$ curve of such a breather 
state displays peculiar features, namely resonant steps and voltage jumps.
The magnitudes of all features are determined by the phase shift between two 
Josephson current oscillations that in turn, can be controlled by an externally 
applied magnetic field.

We thank A. Miroshnichenko, F. Pignatelli, M. Shuster and A. Ustinov for useful 
discussion. This work was supported by the European Union under the RTN 
project LOCNET HPRN-CT-1999-00163.


\begin{thebibliography}{9}
\bibitem{comment}  Eqs. (\ref{beta2}) are not valid in the limit of 
 large discreteness parameter, $\beta_L~\geq~\alpha^{-1}$. However, in this limit 
 the resonant interaction of the breather state with EOs is very weak.
\end{thebibliography}
  \end{document}